\documentclass[conference]{IEEEtran}
\usepackage{amssymb}
\usepackage{amsmath}
\usepackage{txfonts}
\usepackage{graphicx}
\usepackage{multirow}
\usepackage{setspace}
\usepackage{hyperref}
\begin{document}
%\begin{spacing}{2.0}
\bibliographystyle{IEEEtran}
\title{A Selection Region Based Routing Protocol for Random Mobile ad hoc Networks}

\author{\IEEEauthorblockN{Di Li\IEEEauthorrefmark{1}, Changchuan Yin\IEEEauthorrefmark{1}, Changhai Chen\IEEEauthorrefmark{1}, and Shuguang Cui\IEEEauthorrefmark{2}}
\IEEEauthorblockA{\IEEEauthorrefmark{1}Key Laboratory of Universal
Wireless Communications, Ministry of Education\\
Beijing University of Posts and Telecommunications, Beijing, China 100876}
\IEEEauthorblockA{\IEEEauthorrefmark{2}Department of Electrical and Computer Engineering\\
Texas A\&M University, College Station, TX 77843\\
E-mail: dean.lidi@gmail.com, ccyin@ieee.org, chenchanghai@gmail.com, cui@ece.tamu.edu}}

\maketitle

\begin{abstract}
%\boldmath
We propose a selection region based multi-hop routing protocol for random mobile ad hoc networks,
where the selection region is defined by two parameters: a reference distance and a selection
angle. At each hop, a relay is chosen as the nearest node to the transmitter that is located within
the selection region. By assuming that the relay nodes are randomly placed, we derive an upper
bound for the optimum reference distance to maximize the expected density of progress and
investigate the relationship between the optimum selection angle and the optimum reference
distance. We also note that the optimized expected density of progress scales as
$\Theta(\sqrt{\lambda})$, which matches the prior results in the literature. Compared with the
spatial-reuse multi-hop protocol in \cite{Baccelli:Aloha} recently proposed by Baccelli \emph{et
al.}, in our new protocol the amount of nodes involved and the calculation complexity for each
relay selection are reduced significantly, which is attractive for energy-limited wireless ad hoc
networks (e.g., wireless sensor networks).
%We propose a selection region based multi-hop routing protocol for wireless ad hoc networks, where the
%selection region is defined by two parameters, i.e., a reference distance and a selection angle. For each
%hop, a relay is chosen as the nearest node to the transmitter that is located within the reference
%angle and outside of the reference distance. By maximizing the expected density of progress, we
%derive the upper bound on the optimum reference distance and the relationship between the optimum
%selection angle and the optimum reference distance. Further we note that the optimized expected density
%of progress scales as $\Theta(\sqrt{\lambda})$, which conforms to the prior results in the
%literatures. Compared with the spatial reuse multi-hop protocol in \cite{Baccelli-Aloha} recently
%proposed by Baccelli \emph{et al.}, in this new protocol the amount of potential relays involved in
%the relay selection and the calculation complexity for each potential relay are reduced
%significantly, which is attractive for energy-limited wireless ad hoc networks, such as wireless
%sensor networks.
\end{abstract}

% \fi
\IEEEpeerreviewmaketitle
\section{Introduction}
In wireless ad hoc networks, each node may serve as the data source, destination, or relay at
different time instants, which leads to a self-organized network. Such a decentralized structure
makes the traditional network analysis methodology used in centralized wireless networks
inadequate. In addition, it is hard to define and quantify the capacity of large wireless ad hoc
networks. In the seminal work \cite{Capacity:Gupta}, Gupta and Kumar proved that the transport
capacity for wireless
ad hoc networks, defined as the bit-meters pumped every second over a unit area, scales as $\Theta(\sqrt{n})$ in an arbitrary network, where $n$ is node density. %They also
%showed that a typical time-slotted multi-hop architecture with a common transmission range and
%adjacent-neighbor communication can achieve a throughput pre node that scales as
%$\Theta(\frac{1}{\sqrt{n\log n}})$.
In \cite{Weber:Capacity}, Weber \emph{et al.} derived the upper and lower bounds on the
transmission capacity of spread-spectrum wireless ad hoc networks, where the transmission capacity
is defined as the product between the maximum density of successful transmissions and the
corresponding data rate, under a constraint on the outage probability. However, the above work only
considered single-hop transmissions.

In \cite{Sousa:Range}, with multi-hop transmissions and assuming all the transmissions are over the
same transmission range, Sousa and Silvester derived the optimum transmission range to maximize a
capacity metric, called the expected forward progress. Zorzi and Pupolin extended Sousa and
Silvester's work in \cite{Zorzi:RangeFading} to consider Rayleigh fading and shadowing. Recently,
Baccelli \emph{et al.} \cite{Baccelli:Aloha} proposed a spatial-reuse based multi-hop routing
protocol. In their protocol, at each hop, the transmitter selects the best relay so as to maximize
the effective distance towards the destination and thus to maximize the spatial density of
progress. By assuming each transmitter has a sufficient backlog of packets, Weber \emph{et al.} in
\cite{Weber:LongestEdge} proposed longest-edge based routing where each transmitter selects a relay
that makes the transmission edge longest. In \cite{Andrews:RandomAccess}, Andrews \emph{et al.}
defined the random access transport capacity. By assuming that all hops bear the same distance with
deterministically placed relays, they derived the optimum number of hops and an upper bound on the
random access transport capacity.
%indicating that the optimized capacity scales as $\Theta(\sqrt{\lambda})$,
%where $\lambda$ is the density of transmitters

Most of the above works with multi-hop transmissions (e.g., \cite{Sousa:Range},
\cite{Zorzi:RangeFading}, and \cite{Andrews:RandomAccess}) assume that each hop traverses the same
distance, which is not practical when nodes are randomly distributed. On the other hand, in
\cite{Baccelli:Aloha} and \cite{Weber:LongestEdge} the authors proposed routing protocols with
randomly distributed relays; but they did not address how to optimize the transmission distance at
each hop. In this paper, by jointly considering the randomly distributed relays and the
optimization for the hop distance, we propose a selection region based multi-hop routing protocol,
where the selection region is defined by two parameters: a selection angle and a reference
distance. By maximizing the expected density of progress, we derive the upper bound on the optimum
reference distance and the relationship between the optimum reference distance
and the optimum selection angle. %Compared with the routing strategy in \cite{Baccelli-Aloha}, the relay
%region limits the potential relays within a region. Further, since it does not need to compute the
%value of progress towards the destination, the calculation complexity of relay selection for our
%new routing protocol decreases.
%The simulation results show that when the transmission probability is small, the performance of our
%routing protocol tightly closes to that in \cite{Baccelli-Aloha}, which is supposed as the best
%routing strategy with the metric of expected density of progress.
%To our knowledge, this is the first work to derive the optimum reference distance for multi-hop
%networks with the consideration that relay nodes are random distributed, while the previous
%literatures \cite{Sousa-Opt-range}, \cite{Zorzi-Optimum-range-with-fading} and
%\cite{Random-Access-Transport-Capacity} usually assumed that relay nodes are predetermined located
%along the line from the source to the destination.

The rest of the paper is organized as follows. The system model and the routing protocol are
described in Section II. The selection region optimization is presented in Section III. Numerical
results and discussions are given in Section IV. The computational complexity is analyzed in
Section V. Finally, Section VI summarizes our conclusions.
\section{System Model and Routing Protocol}
In this section, we first define the network model, then present the selection region based routing
protocol.
\subsection{Network Model}
Assume nodes in the network follow a homogenous Poisson Point Process (PPP) with density $\lambda$,
with slotted ALOHA being deployed as the medium access control (MAC) protocol. We also consider the
nodes are mobile, to eliminate the spatial correlation, which is also discussed in
\cite{Baccelli:Aloha}. During each time slot a node chooses to transmit data with probability $p$,
and to receive data with probability $1-p$. Therefore, at a certain time instant, transmitters in
the network follow a homogeneous PPP ($\Pi_{Tx}$) with density $p\lambda$, while receivers follow
another homogenous PPP ($\Pi_{Rx}$) with density $(1-p)\lambda$. Considering multi-hop
transmissions, at each hop a transmitter tries to find a receiver in $\Pi_{Rx}$ as the relay. We
assume that all transmitters use the same transmission power $\rho$ and the wireless channel
combines the large-scale path-loss and small-scale Rayleigh fading. The normalized channel power
gain over distance $d$ is given by
\begin{equation}
G(d) = \frac{\gamma }{{{d^\alpha }}},
\end{equation}
where $\gamma$ denotes the small-scale fading, drawn from an exponential distribution of mean
$\frac{1}{\mu}$ with probability density function (PDF) $f_\gamma(x)=\mu \exp(-\mu x)$, and
$\alpha>2$ is the path-loss exponent.

For the transmission from transmitter $i$ to receiver $j$, it is successful if the received
signal-to-interference-plus-noise ratio (SINR) at receiver $j$ is above a threshold $\beta$. Thus
the successful transmission probability over this hop with distance $d_{ij}$ is given by
\begin{equation}
P_s=\Pr\left(\frac{\rho\gamma_0d_{ij}^{-\alpha}}{\sum_{k\in\Pi_{TX}\setminus\{i\}}\rho\gamma_i{d_{kj}}^{-\alpha}+\eta}>\beta\right),
\end{equation}
where $i\in \Pi_{Tx}$, $j\in \Pi_{Rx}$,
$\sum_{k\in\Pi_{Tx}\setminus\{i\}}\rho\gamma_i{d_{kj}}^{-\alpha}$ is the sum interference from the
simultaneous concurrent transmissions, $d_{kj}$ is the distance from interferer $k$ to receiver
$j$, and $\eta$ is the average power of ambient thermal noise. In the sequel we approximate
$\eta=0$, which is reasonable in interference-limited ad hoc networks. From \cite{Baccelli:Aloha},
the successful transmission probability from transmitter $i$ to receiver $j$ is derived as
\begin{equation}
    P_s= \exp\left(-\lambda{p}t{d_{ij}}^{2}\right),
\end{equation}
where \[t = \frac{{2{\pi ^2}/\alpha }}{{\sin (2\pi /\alpha )}}{\beta ^{2/\alpha}}.\tag{3a}\]
\subsection{Selection Region Based Routing}
%For multi-hop network, if we assume position-determined relays exist to ensure each hop shares the
%same distance that aggregates to form the path from the data source to its final destination, the
%optimum transmission distance for one hop can be derived. In this condition we actually use the one
%hop distance to determine the location of relays, i.e., deterministically located relays, but in ad
%hoc network terminals are usually random distributed.
Considering a typical multi-hop transmission scenario, where a data source (S) sends information to
its final destination (D) that is located far away, and it is impossible to complete this operation
over a single hop.% Thus a multi-hop transmission is needed. In multi-hop wireless networks, if we assume
%position predetermined relays exist equidistantly along the line from the source to the
%destination, which means that each hop
%shares the same transmission distance as shown in Fig. 1, %and each hop shares
%%the same distance which aggregates to form the path from the data source to its final destination,
%the optimum transmission distance for each hop can be derived in
%\cite{Random-Access-Transport-Capacity}. %In this case, actually the optimum transmission distance
%is used to determine the location of a relay.

Since we assume that nodes are randomly distributed, relays may not be located at an optimum
transmission distance as derived in \cite{Andrews:RandomAccess}. To guarantee a relay existing at a
proper position, we propose a selection region based multi-hop routing protocol. For each
transmitter along the route to the final destination, we define a selection region by two
parameters: a selection angle $\varphi$ and a reference distance $r_m$, as shown by the shaded area
in Fig. 1, where the selection region is defined as the region that is located within angle
$\varphi$ and outside the arc $\widehat{AB}$ with $|\overrightarrow{OB}|=r_m$. Here, the
transmitter is placed in the circle center $O$, $\angle {BOC}=\angle {AOC}=\varphi/2$, and
$\overrightarrow{OC}$ points to the direction of the final destination. At each hop, the relay is
selected as the nearest receiver node to the transmitter among the nodes in the selection region.

The reason that we limit the selection region within an angle $\varphi$ is explained as follows: In
multi-hop routing, a transmission is inefficient if the projection of transmission distance on the
directional line from the transmitter towards the final destination is negative, or less efficient
if the projection is positive but very small. Therefore, here we set a limiting angle $\phi$ with
which each packet traverses at each hop within $[-\varphi/2,\varphi/2]$.
\begin{figure}
\begin{center}
\includegraphics[width=0.35\textwidth]{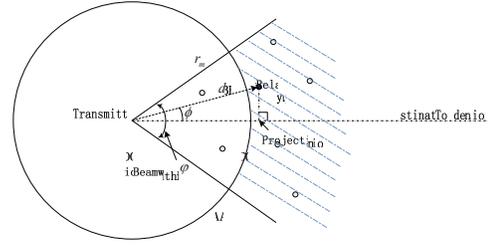}
\caption{Selection region} \label{Scenario}
\end{center}
\end{figure}

Compared with the model in \cite{Andrews:RandomAccess}, where the authors assume that the relays
are equidistantly placed along a line from the source to the destination as shown in Fig. 2, here
our model assumes that the intermediate relays are randomly distributed over the selection region
following a homogenous PPP ($\Pi_{Rx}$) with density $(1-p)\lambda$, which is more practical.
\begin{figure}
\begin{center}
\includegraphics[width=0.3\textwidth]{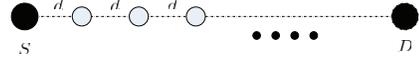}
\caption{Ideal scenario: position-predetermined relays are equidistantly placed along the line from
the source to the destination} \label{DeterminedScenario}
\end{center}
\end{figure}
\section{Selection Region Optimization}
In this section, we optimize the selection region to derive the optimum values for the selection
angle $\varphi$ and the reference distance $r_m$ by maximizing the expected density of progress.

As in \cite{Baccelli:Aloha}, the density of progress is defined as
\[D=p\lambda  \cdot P_{s} \cdot d \cos \phi,\]
where $P_s$ is the successful transmission probability defined in (2), $d \cos \phi$ is the
projection of the transmission distance $d$ along the directional line $\overrightarrow{OC}$. Since
the receivers follow a homogeneous PPP with density $\lambda(1-p)$, the cumulative distribution
function (CDF) of the transmission distance $d$ is given as
\begin{equation}
    \Pr (d \leq r) = 1- \exp \left[ - \lambda (1 - p)\frac{\varphi }{2}({r^2} - r_m^2)\right],{r_m} \le r <
    \infty.
\end{equation}
Since $\phi$ is uniformly distributed over $[-\varphi/2,\varphi/2]$, which is independent of $d$,
the expected density of progress is given by
\begin{align}\label{E(d)}
 E[D] &= p\lambda \int\limits_{r_m}^\infty \int\limits_{-\frac{\varphi}{2}}^{\frac{\varphi}{2}}  {{e^{ - p\lambda t{x^2}}}} x \cos \phi {f_{d}}(x)d\phi dx  \nonumber\\
  &= \sqrt \lambda  p(1 - p)\Gamma \left(\frac{3}{2},kr_m^2\right){k^{ - 3/2}}\exp \left(\lambda (1 - p)\frac{\varphi }{2}r_m^2\right)\sin \left(\frac{\varphi
  }{2}\right),
\end{align}
where ${f_{d}}(x)$ is the PDF of $d$ obtained from (4), $k = pt + (1 - p)\frac{\varphi }{2}$, $t$
is defined in (3a), and $\Gamma \left(\frac{3}{2},kr_m^2\right) = \int\limits_{kr_m^2}^\infty {{e^{
- t}}{t^{\frac{3}{2} - 1}}dt} $ is the incomplete Gamma function.

To optimize the objective function in (5), we first assume that $\varphi$ is constant, and try to
derive the optimum value of $r_m$. For brevity, in the following derivation we write the objective
function as $E$. Setting the derivative with respect to $r_m$ as 0, after some calculations we have
\begin{align}\label{E/rm}
\frac{{dE}}{{dr_m^{}}} = &\exp \left(\lambda (1 - p)\frac{\varphi }{2}r_m^2\right)\nonumber\\
&\cdot\left[\Gamma \left(\frac{3}{2},kr_m^2\right)\lambda (1 - p)\varphi r_m^{} +
\frac{{d\Gamma\left(\frac{3}{2},kr_m^2\right)}}{{dr_m^{}}}\right] = 0,
\end{align}
where $\Gamma \left(\frac{3}{2},kr_m^2\right)$ is calculated as
\begin{equation}
\Gamma \left(\frac{3}{2},kr_m^2\right) = \Gamma \left(\frac{3}{2}\right) + \sqrt k {r_m}\exp \left(
- kr_m^2\right) - \frac{{\sqrt \pi  }}{2}erf\left(\sqrt k {r_m}\right).
\end{equation}

Thus we have
\begin{align}
 \frac{{d\Gamma\left(\frac{3}{2},kr_m^2\right)}}{{dr_m^{}}} & =  \sqrt k \exp \left( - kr_m^2\right)\left(1 - 2kr_m^2\right) - \sqrt k \exp \left( - kr_m^2\right) \nonumber\\
& = - 2{k^{3/2}}r_m^2\exp \left(- kr_m^2\right).
 \end{align}

Applying (8) to (6), we obtain
\begin{equation}
    \Gamma\left(\frac{3}{2},kr_m^2\right)\lambda \left(1 - p\right)\varphi {r_m} - 2{k^{3/2}}r_m^2\exp \left(- kr_m^2\right) = 0.
\end{equation}

Note that the above is only the necessary condition for optimality, given the unknown convexity of
objective function. However, the global optimum must be among all the roots of the above equation,
which can be found numerically. Since it is difficult to analytically derive the exact solution for
$r_m$ from (9), we turn to get an upper bound of $r_m$. Since
\begin{align}\label{Gamma}
    \Gamma \left(\frac{3}{2},kr_m^2\right) &> \frac{1}{2}\left[\Gamma \left(1,kr_m^2\right) + \Gamma \left(2,kr_m^2\right)\right] \nonumber\\
    &= \frac{1}{2}\exp \left( - kr_m^2\right)\left(2 + kr_m^2\right),
\end{align}
by (9), we have
\begin{equation}
    \lambda (1 - p)k\varphi r_m^2 - 4{k^{3/2}}r_m^{} + \lambda (1 - p)\varphi  > 0.
\end{equation}
Therefore,
\begin{equation}
    {r_m} < \frac{{2{k^{3/2}} - \sqrt {4{k^3} - 2k{{[\lambda (1 - p)\varphi ]}^2}} }}{{k\lambda (1 - p)\varphi
    }}.
\end{equation}
\begin{figure}
\begin{center}
\includegraphics[width=0.4\textwidth]{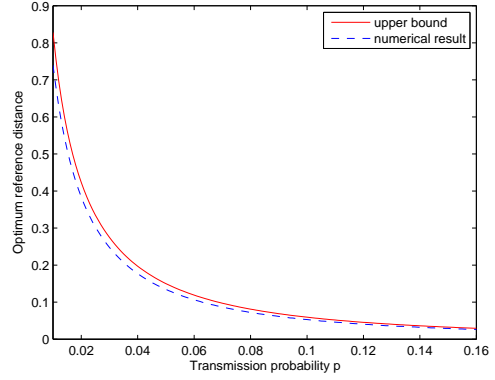}
\caption{Numerical results and the analytical upper bound for the optimum reference distance}
\label{2}
\end{center}
\end{figure}

In Fig. 3, we compare the upper bound of the optimum $r_m$ with the numerically computed optimal
value when $\varphi=\frac{1}{3}\pi$. We see that when transmission probability $p$ increases, the
upper bound becomes tighter.

Now let's maximize the objective function by jointly optimizing $r_m$ and $\varphi$. Rewrite (5) as
\[E = \sqrt \lambda  p(1 - p)\exp \left( - \lambda ptr_m^2\right)\Gamma \left(\frac{3}{2},kr_m^2\right)\exp \left(\lambda
kr_m^2\right){k^{ - 3/2}}\sin \left(\frac{\varphi }{2}\right).\] For brevity, let us denote $\exp
(\lambda kr_m^2)$ as $e$ and $\Gamma \left(\frac{3}{2},kr_m^2\right)$ as $\Gamma$. With partial
derivatives, we have
\begin{align}
\frac{{\partial E}}{{\partial r_m^{}}}=&\sqrt \lambda  p(1 - p)\sin \left(\frac{\varphi
}{2}\right){k^{ -
3/2}}\exp \left( - \lambda ptr_m^2\right)\nonumber\\
&\cdot \left[ - 2\lambda pt{r_m}\Gamma e + \Gamma \frac{{\partial e}}{{\partial r_m^{}}} +
e\frac{{\partial \Gamma }}{{\partial r_m^{}}}\right] = 0.
\end{align}

This holds only if
\begin{equation}
\Gamma \frac{{\partial e}}{{\partial r_m^{}}} + e\frac{{\partial \Gamma }}{{\partial
r_m^{}}}=2\lambda pt{r_m}\Gamma e.
\end{equation}

Since $k = pt + (1 - p)\frac{\varphi }{2}$, there is $\frac{{\partial k}}{{\partial \varphi }} =
\frac{1}{2}(1 - p)$. To simplify things, we can then calculate the derivative with respect to $k$
instead of $\varphi$ as
\begin{flalign}
\frac{{\partial E}}{{\partial k}} &= \sqrt \lambda p{(1 - p)}\exp\left( - \lambda ptr_m^2\right)\nonumber\\
&\cdot \left[(\frac{{\cos (\varphi /2)}}{{1 - p}} - \frac{3}{2}\sin \left(\frac{\varphi
}{2}\right){k^{ - 1}})\Gamma e + \sin \left(\frac{\varphi }{2}\right)\left(\Gamma \frac{{\partial
e}}{{\partial k}} + e\frac{{\partial \Gamma }}{{\partial k}}\right)\right] = 0.
\end{flalign}

Since the factor in $e$ and $\Gamma$ related to $k$ and $r_m$ is only $kr_m^2$, thus we get
$\frac{1}{2}r_m^{}\frac{{\partial e}}{{\partial r_m^{}}} = k\frac{{\partial e}}{{\partial k}}$ and
$\frac{1}{2}r_m^{}\frac{{\partial \Gamma }}{{\partial r_m^{}}} = k\frac{{\partial \Gamma
}}{{\partial k}}$.

Therefore, with (14), we have
\begin{equation}
\Gamma \frac{{\partial e}}{{\partial k}} + e\frac{{\partial \Gamma }}{{\partial k}} =
\frac{{{r_m}}}{{2k}}\left(\Gamma \frac{{\partial e}}{{\partial {r_m}}} + e\frac{{\partial \Gamma
}}{{\partial {r_m}}}\right) = {k^{ - 1}}\lambda ptr_m^2\Gamma e.
\end{equation}

Applying (16) to (15), the following holds:
\begin{equation}
\left[\frac{{\cos (\varphi /2)}}{{1 - p}} - \frac{3}{2}\sin \left(\frac{\varphi }{2}\right){k^{ -
1}}\right]\Gamma e + \sin \left(\frac{\varphi }{2}\right){k^{ - 1}}\lambda ptr_m^2\Gamma e = 0.
\end{equation}

After some calculation, (17) is simplified as
\begin{equation}
\frac{{\cot (\varphi /2)k}}{{1 - p}} - \frac{3}{2} + \lambda ptr_m^2 = 0.
\end{equation}

Since it is hard to derive close-formed solutions for the optimal $r_m$ and $\varphi$,
respectively, we implicitly use $\varphi$ and $p$ to express the optimal $r_m$ as

 \begin{equation}\label{Solution}
 r_m^{} = \sqrt {\frac{3/2-{\cot (\varphi /2)k/(1 - p)}}{{\lambda pt}}}.
 \end{equation}

Note that $r_m$ scales as $\lambda^{-1/2}$, which intuitively makes sense. Since as the density
increases, the interferers' relative distance to the receiver decreases as $\sqrt{\lambda}$, it
requires a shorter transmission distance by the same amount to keep the required SINR. By applying
(19) in (5), we observe that (5) becomes $N\sqrt{\lambda}$, where $N$ is a constant independent of
$\lambda$. This means that the maximum expected density of progress scales as
$\Theta(\sqrt{\lambda})$, which conforms to the results in \cite{Capacity:Gupta} and
\cite{Andrews:RandomAccess}.
\section{Numerical Results}
In this section, we present some numerical results based on the analysis in Section III. We choose
the path-loss exponent $\alpha$ as 3, the node density $\lambda$ as 1, and the outage threshold
$\beta$ as 10 dB. In Figs. 4 and 5, we plot the expected density of progress vs. the reference
distance $r_m$ and the selection angle $\varphi$, with $p=0.01$ and 0.05, respectively. We see that
for each $p$ there exists an optimum selection angle and an optimum reference range when the
respective partial derivatives are zero as discussed in Section III.
\begin{figure}
\begin{center}
\includegraphics[width=0.4\textwidth]{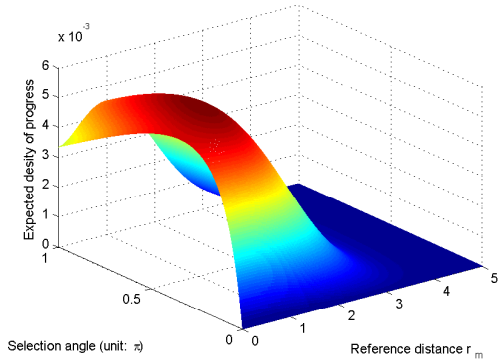}
\caption{The expected density of progress vs. the reference distance and the selection angle with
p=0.01} \label{3}
\end{center}
\end{figure}
\begin{figure}
\begin{center}
\includegraphics[width=0.4\textwidth]{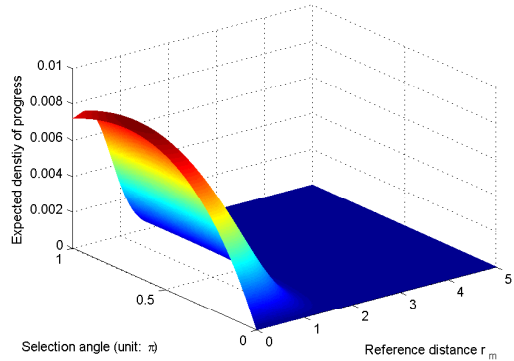}
\caption{The expected density of progress vs. the reference distance and the selection angle with
p=0.05} \label{4}
\end{center}
\end{figure}

In Fig. 6, we plot the optimum selection angle obtained numerically vs. the transmission
probability. As shown in the figure, we see that the increment of transmission probability leads to
the increase of the optimum selection angle. This can be explained as follows: The increment of
transmission probability means the decrement of the number of nodes that can be selected as relays;
therefore the selection angle should be enlarged to extend the selection region.
\begin{figure}
\begin{center}
\includegraphics[width=0.4\textwidth]{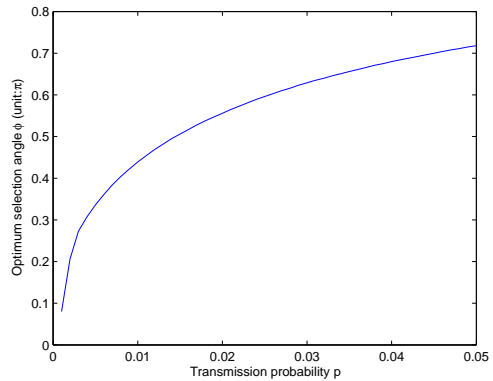}
\caption{The optimum selection angle vs. the transmission probability p} \label{5}
\end{center}
\end{figure}

In Fig. 7, we compare the optimum reference distance obtained numerically with that derived in
(19), where $\varphi$ is chosen optimally as that in Fig. 6. We see that the increment of
transmission probability leads to the decease of reference distance. This can be explained as
follows: The increment of transmission probability means more simultaneous concurrent transmissions
such that the interference will be increased; therefore the reference transmission distance should
be decreased to guarantee the quality of the received signal and the probability of successful
transmission.

In Fig. \ref{7}, we compare the performance of our routing protocol with that in
\cite{Baccelli:Aloha}, and also with the optimized (with optimum angle) nearest neighbor routing
shown in Fig. 5 of \cite{Baccelli:Aloha} and the non-optimized (with an arbitrary angle, e.g.,
$\frac{\pi}{2}$) nearest neighbor routing in \cite{Martin:Routing}.
\begin{figure}
\begin{center}
\includegraphics[width=0.4\textwidth]{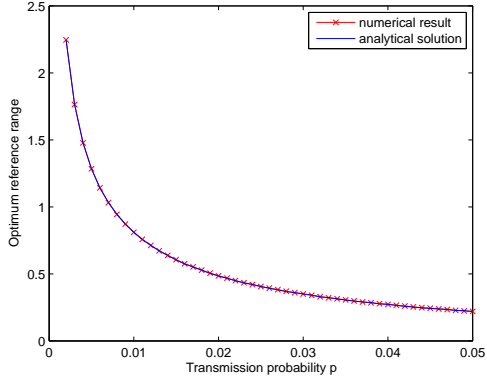}
\caption{The optimum reference distance vs. the transmission probability p} \label{6}
\end{center}
\end{figure}
\begin{figure}
\begin{center}
\includegraphics[width=0.4\textwidth]{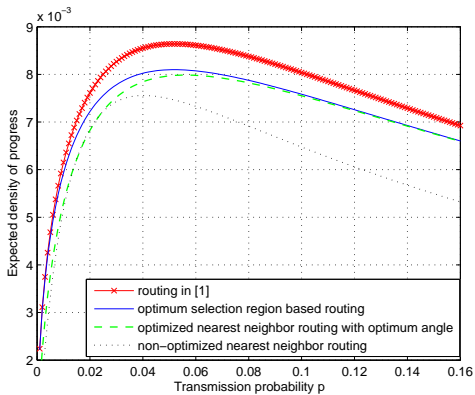}
\caption{The performance of different routing protocols} \label{7}
\end{center}
\end{figure}

From Fig. 8, we have the following observations and interpretations: 1)When $p$ increases, the
performance of our routing protocol becomes close to that of the nearest neighbor routing with an
optimum angle. This can be explained from Fig. 7 as: When $p$ increases, the reference distance
$r_m$ tends to be a small value close to zero; thus our routing scheme degenerates to the nearest
neighbor routing. Furthermore, our routing protocol shows much better performance than the
non-optimized (with non-optimized angle) nearest neighbor routing, and this advantage is due to
adopting both the optimum selection angle and the optimum reference distance. In this case, the
selection region based routing can also be considered as the optimized nearest neighbor routing
given the selection angle and the reference distance; 2) We see that when at the optimum selection
angle and the optimum reference distance, the optimum transmission probability is approximately
0.05. Although in this paper we mainly focus on the optimization of the selection region, this
observation indicates that there also exists an optimum transmission probability with our model,
which has been discussed in some other prior literature, e.g., \cite{Baccelli:Aloha},
\cite{Sousa:Range}, \cite{Zorzi:RangeFading}, and \cite{Weber:LongestEdge}.

\section{Discussion on Complexity}
As shown in Fig. 8, for multi-hop ad hoc networks, in terms of the performance metric of expected
density of process, Baccelli \emph{et al}.'s routing strategy is the best, by design at each hop
the transmitter chooses a relay that provides the maximum value of progress towards the
destination. However, the computational complexity with this protocol might be high, since the
transmitter at each hop should compute the successful transmission probability $P_{s|x|}$ together
with the projection of transmission distance, and accordingly evaluate the value of progress
towards the destination for each receiver, further choose the one with the greatest value of
progress as the relay.

In our protocol as we see from Fig. 8, when $p$ is small, its performance is close to that of
Baccelli \emph{et al.}, while the computational complexity per hop is reduced significantly:

1) The nodes involved in the relay selection process are limited to a small region. As relays are
selected from the receiver nodes in the selection region, the number of nodes participating in the
relay selection is reduced with a ratio $\varphi/{2\pi}$ compared with that in
\cite{Baccelli:Aloha}.

2) Unlike that in \cite{Baccelli:Aloha}, where the successful transmission probability $P_{s|x|}$,
the projection of transmission distance, and further the value of progress towards the destination
for each potential relay need to be calculated; we only need to calculate the distance between the
transmitter and the potential relays.

Also note that our new protocol could be easily implemented by deploying directional antennas in
the transmitter, where the spread angle can be set equal to the optimum selection angle. In this
case not only the network computational complexity but also the interference will be reduced, which
will be addressed in our future work.
\section{Conclusions}
In this paper, we propose a selection region based multi-hop routing protocol for random mobile ad
hoc networks, where the selection region is defined by two parameters, a selection angle and a
reference distance. By maximizing the expected density of progress, we present some analytical
results on how to refine the selection region. Compared with the previous results in
\cite{Sousa:Range}, \cite{Zorzi:RangeFading}, and \cite{Andrews:RandomAccess}, we consider the
transmission direction at each hop towards the final destination to guarantee relay efficiency.
Compared with the protocol in \cite{Baccelli:Aloha}, the optimum selection region defined in this
paper limits the area in which the relay is being selected, and the routing computational
complexity at each hop is reduced.


%\bibliographystyle{IEEEtran}
%\bibliography{IEEEbrv,../bib/paper}
%\begin{thebibliography}{1}

%\bibliography{Optimization}

%\end{spacing}

%\cite{}
\end{document}